\documentclass[11pt]{article}

\usepackage[english]{babel}
\usepackage{amsmath}
\usepackage{amssymb}
\usepackage{graphicx}
\usepackage{subcaption}
\usepackage{hyperref}
\usepackage{placeins}
\usepackage{siunitx}
\usepackage{authblk}

\newcommand{\disp}{\mathbf{u}}
\newcommand{\eps}{\boldsymbol{\varepsilon}(\mathbf{u})}

\newcommand{\tr}{\text{tr}}

\usepackage[a4paper,top=2cm,bottom=2cm,left=2cm,right=2cm,marginparwidth=1.75cm]{geometry}

\usepackage{amsmath}
\usepackage{graphicx}

\usepackage{csquotes}
\usepackage[backend=biber]{biblatex}
\title{Origins of phase-field crack widening in dynamic fragmentation explained}
\author[1]{Shad Durussel\footnote{Corresponding author: shad.durussel@epfl.ch}}
\author[2]{Gergely Molnár}
\author[1]{Jean-François Molinari}

\affil[1]{\small \textit{Institute of Civil Engineering, Institute of Materials Science and Engineering, École Polytechnique Fédérale de Lausanne (EPFL), 1015 Lausanne, Switzerland}}
\affil[2]{\small \textit{CNRS, INSA Lyon, LaMCoS, UMR5259, 69621 Villeurbanne, France}}

\date{}

\begin{document}
\maketitle

\begin{abstract}

We investigate dynamic crack propagation and fragmentation with the phase-field fracture approach. The method was chosen for its ability to yield crack paths that are independent of the underlying mesh, thanks to the damage regularization zone. In dynamics, we observe a progressive widening of this regularization zone and attribute it to an unphysical trapping of elastic waves. We show that the damage zones do not represent free boundaries accurately and that wave interactions induce additional damage. 
We reveal how mass erosion, by conserving the elastic wave speed in the damaged regions, can be used to efficiently reduce the spurious diffusion of damage. Furthermore, we provide numerical evidence that dynamically propagating cracks in the phase-field formulation, both with and without mass erosion, converge to the predictions of linear elastic fracture mechanics. For vanishing regularization length, the crack speed and energy release rate become independent of the phase-field regularization length, provided that this length scale is small enough and the mesh fine enough to resolve the process zone. 

\textbf{Keywords: } phase-field fracture; dynamic fragmentation; mass erosion; branching; wave-damage interaction

\end{abstract}
 
\section{Introduction}

Dynamic fragmentation refers to the rapid breakup of a solid into multiple fragments resulting from the sudden release of a large amount of energy, such as during an impact or an explosion. 
The study of this phenomenon has a wide range of applications, from improving the safety of tempered glass~\cite{acloque1956deferred, vocialta_numerical_2018, corrado2024phase} to mitigating the risks associated with space debris~\cite{krisko2008socit4, morgado2022multi}. 

Theoretical fragmentation models were developed to predict fragment size distributions under dynamic loading, initially using a statistical approach~\cite{mott_theory_1943} and later extended to energy-driven models~\cite{grady_fragmentation_1983, glenn_chudnovsky_1986, ramesh2015review}. The complexity and variability of real-world scenarios, however, motivate the use of computational models that can accommodate more complex geometries, loading conditions, and material behaviour. The end goal is to accurately capture the elastodynamics as well as the fracture mechanics leading to crack nucleation and propagation (including branching), which collectively shape the final fragments size distribution.

Discrete element simulations have demonstrated that even relatively simple particle models can reproduce key statistical features of fragmentation, such as power-law fragment size distributions following impacts or internal explosions in solid bodies~\cite{kun1996study, wittel2004fragmentation}.
In contrast, mesh-based numerical fracture models are able to describe the formation and evolution of individual cracks that ultimately lead to fragmentation. 
These fracture models can be broadly classified into two main categories.

The first category comprises methods based on a discrete representation of discontinuities. 
For instance, the extended finite element method (XFEM) enriches the interpolation space to allow for the discrete representation of cracks without remeshing~\cite{moes1999finite}, but requires additional criteria to handle crack branching~\cite{belytschko2003dynamic}. Cohesive zone models (CZM) provide an explicit representation of cracks and fragments through topological changes in the mesh~\cite{dugdale1960yielding, barenblatt1962mathematical}, either along predefined crack paths~\cite{xu1994numerical} or via the automatic insertion of cohesive elements~\cite{camacho_computational_1996}. 
This makes CZM particularly well suited for the study of fragmentation~\cite{ortiz1999finite, ZHOU20055181,vocialta_3d_2017, vocialta_numerical_2018}. However, without remeshing, this method remains mesh dependent, which limits the robustness of fracture path predictions and, consequently, of the resulting fragment shapes.

The second category includes models based on a smeared representation of cracks. 
Representative examples include nonlocal integral approaches~\cite{pijaudier1987nonlocal, jirasek1998nonlocal}, peridynamics~\cite{bobaru2015cracks}, thick level-set methods~\cite{moes2011level}, and the Lip-field approach~\cite{chevaugeon2022lipschitz} amongst others. 
Phase-field modelling of fracture also falls within this category. 
As with other smeared approaches, phase-field methods are attractive for their independence from the underlying mesh topology, but they require sufficiently fine mesh resolution to accurately capture the transition from intact to fully broken material. Phase-field fracture models were developed within both the physics community, through Ginzburg–Landau-type phase-transition formulations~\cite{karma_phase-field_2001, henry_study_2008}, and the mechanics community, through variational formulations of brittle fracture based on Griffith’s theory~\cite{griffith1921griffith, francfort_revisiting_1998} and their regularized approximations~\cite{ambrosio_approximation_1990, bourdin_numerical_2000}. The crack thickness in phase-field models can be interpreted in two ways: as a purely numerical regularization that should vanish as the regularization length is refined, or as a genuine material parameter~\cite{molnar2020toughness}.
This approach has since been adapted to a wide range of fracture applications, including ductile fracture~\cite{ambati2015phase} and stochastic modelling of brittle fracture~\cite{gerasimov_stochastic_2020}.

The extension of the phase-field framework to dynamic crack propagation~\cite{borden_phase-field_2012, li_gradient_2016} has opened the door to its application to dynamic fragmentation, and it has already been used to study one-dimensional fragmentation~\cite{geromel2019gradient, HUYNH2025105971}. 
However, a widening of damage bands is commonly observed under dynamic loading~\cite{borden_phase-field_2012, li_gradient_2016}. 
As this widening appears to be load-dependent, it becomes particularly problematic in the context of dynamic fragmentation, where extreme loading can lead to uninterpretable and potentially unphysical damage patterns. Moreover, because the damaged material is progressively degraded, elastic waves are reflected by a diffusive interface, potentially leading to additional damage and spalling through wave superposition~\cite{WEINBERG2022114330}. 
This suggests that the interaction between elastic waves and damaged zones is a key mechanism in the development of this damage sprawl.

As material stiffness is degraded, the wave speed within the damaged zone decreases, and elastic waves interact with damage. 
The objective of this work is to compare the classical phase-field formulation with an alternative approach that preserves the wave velocity by degrading the mass consistently with the stiffness. 
After illustrating the issue in a fragmentation example, we investigate the interaction between an elastic wave and a damaged boundary in a pseudo one-dimensional setting and explore, how damage spreads through this interaction.

The differences in crack propagation dynamics between the two approaches are then examined in two dimensions, first for a single crack and then in the presence of branching. 
In this context, we study the convergence of the crack velocity and the dissipated energy as the regularization length is reduced.

This article begins by recalling the formulation of phase-field modelling of fracture in Section~\ref{sec:method}. 
In Section~\ref{sec:fragmentation_full}, we illustrate the issue of damage-band widening in fragmentation. 
The interaction between elastic waves and damage, as well as the resulting damage sprawl, is then explored in Section~\ref{sec:1d_bar}.
In Section~\ref{sec:2d}, we investigate the behaviour of the model with mass degradation in the context of two-dimensional crack propagation and branching. 
Finally, Section~\ref{sec:conclusion} discusses the perspectives opened by these results and outlines potential directions for future research.

\FloatBarrier
\section{Phase-field modelling of brittle fracture}
\label{sec:method}

\subsection{Regularized variational formulation of Griffith energy balance}

Starting from the free-discontinuity framework, brittle fracture can be described by the variational formulation introduced by Francfort and Marigo~\cite{francfort_revisiting_1998}.
In this setting, the displacement field $\disp$ and the crack set $\Gamma$ are obtained by minimizing the total energy functional
\begin{equation}
  E(\disp, \Gamma) = \int_{\Omega \setminus \Gamma} \Psi (\eps) \, dV
  + G_c \mathcal{H}^{n-1}(\Gamma),
  \label{eq:variational_griffith}
\end{equation}
where $\Omega$ denotes the domain of interest and $\Gamma$ represents a set of discontinuities corresponding to cracks.
The first term accounts for the elastic strain energy stored in the intact region $\Omega \setminus \Gamma$, with $\Psi(\varepsilon)$ the strain-energy density expressed in terms of the linearized strain tensor $\varepsilon(\mathbf{u})$.
The second term represents the fracture energy required to create new crack surfaces. It is proportional to the critical energy release rate $G_c$, and to $\mathcal{H}^{n-1}(\Gamma)$, the $(n-1)$-dimensional Hausdorff measure quantifying the total crack surface.

\begin{figure}[ht]
  \centering
  \begin{subfigure}{0.58\linewidth}
    \centering
    \includegraphics{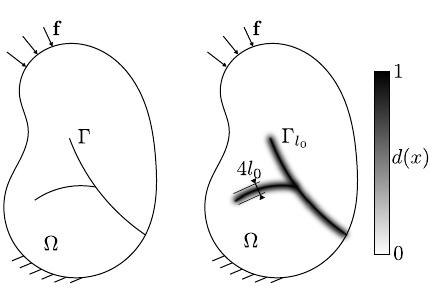}
    \caption{}
    \label{fig:phasefield_model}
  \end{subfigure}
  \begin{subfigure}{0.4\linewidth}
  	\centering
    \includegraphics{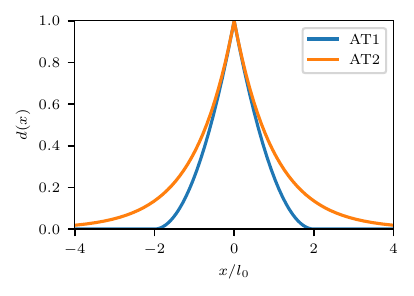}
    \caption{}
    \label{fig:phasefield_profile}
  \end{subfigure}	
  \captionsetup{subrefformat=parens}
    \caption{\subref{fig:phasefield_model} Regularization of a sharp crack $\Gamma$ by a diffusive damage field $d(\mathbf{x})$ in a continuous domain $\Omega$ subject to traction $\mathbf{f}$ and imposed displacement. The regularized fracture corresponds to the region $\Gamma_{l_0} = \{\mathbf{x} \in \Omega | d(\mathbf{x}) > 0\}$ of width related to $l_0$. \\
    \subref{fig:phasefield_profile} Optimal damage profile for a 1D crack located at $x=0$. Note the limited support of the damage for the AT1 formulation.}
    \label{fig:phasefield_method}
\end{figure}	
  
Regularization of the free-discontinuity problem follows the framework introduced in Ref.~\cite{ambrosio_approximation_1990}.
In particular, the work of Bourdin et al.~\cite{bourdin_numerical_2000} proposes a variational regularization based on a scalar damage field $d(\mathbf{x}) \in [0,1]$, where $d = 0$ denotes undamaged material, $d = 1$ corresponds to fully damaged material, and intermediate values describe a diffusive transition zone.
This transition from a sharp discontinuity to a smeared description of the crack is illustrated in Figure~\ref{fig:phasefield_model}.
The regularized energy functional is given by
\begin{equation}
	E_{l_0}(\disp, d) = \int_{\Omega} g(d)\Psi (\eps) \, dV
	+ \frac{G_c}{c_w}\int_{\Omega} \left( \frac{w(d)}{l_0} + l_0 \|\nabla d\|^2 \right) \, dV,
  \label{eq:phase-field_regularization}
\end{equation}
where the Hausdorff measure in Equation~(\ref{eq:variational_griffith}) is approximated by a volumetric integral.
The length scale parameter $l_0$ governs the thickness of the diffusive crack region, i.e. the width of the transition zone between $d=0$ and $d=1$.
The degradation function $g(d)$ is a continuous, monotonically decreasing function that reduces the material stiffness as damage grows, and satisfies $g(0)=1$, $g(1)=0$, $g'(1)=0$ and $g'(d) < 0$ for $d \in [0,1)$.
A widely used choice for this function is the quadratic polynomial
\begin{equation}
  g(d) = (1 - d)^2.
\end{equation}

The local contribution to the dissipated energy is described by a continuous and monotonic function $w(d)$ satisfying $w(0)=0$, $w(1)=1$ and $w'(d) \geq 0$ for $d \in [0,1]$.
This function is associated with a normalization constant defined as $c_w := 4\int_0^1 \sqrt{w(s)} ds$.

The original formulation in~\cite{bourdin_numerical_2000} gives $w(d) = d^2$ and $c_w = 2$.
In~\cite{pham_gradient_2011}, a linear function is proposed for the dissipated energy.
These two formulations are commonly referred to as AT2 and AT1, respectively:
\begin{itemize}
    \item AT1: \quad $w(d) = d$, \quad $c_w = \frac{8}{3}$
    \item AT2: \quad $w(d) = d^2$, \quad $c_w = 2$
\end{itemize}
The optimal damage profile for a one-dimensional crack located at $x=0$ is shown in Figure~\ref{fig:phasefield_profile} for both the AT1 and AT2 formulations.

The main difference between the two models is that AT1 predicts an initial purely elastic phase prior to damage nucleation, whereas AT2 produces a gradual degradation of stiffness from the onset of loading.
The inherent positivity of the damage variable in AT2, resulting from the squared term in the dissipation function, is often considered an attractive feature.
However, in dynamic applications it is desirable to preserve the undamaged elastic response and the correct wave speed away from the crack, which makes the AT1 formulation more suitable in that context.

The problem is extended to dynamic by accounting the contribution of the kinetic energy to the total energy functional. Subtracting also the work from external forces $W_{\text{ext}}(\disp)$ gives
\begin{equation}
	E_{l_0}(\disp, d) = \int_{\Omega} g(d)\Psi (\eps) \, dV
	+ \frac{G_c}{c_w}\int_{\Omega} \left( \frac{w(d)}{l_0} + l_0 \|\nabla d\|^2 \right) \, dV
    - \int_{\Omega} \frac{1}{2} \rho \dot{\disp} \cdot \dot{\disp}\, dV
    - W_{\text{ext}}(\disp),
  \label{eq:phase-field_dynamic}
\end{equation}
with $\rho$ the mass density of the material.

The solution to the problem is obtained by minimizing the total energy with the displacement $\mathbf{u}$ and damage $d$ as unknowns:
\begin{equation}
  \min_{\mathbf{u}, d\geq d_{n-1}} E_{l_0}
\end{equation}

In practice, a staggered scheme is often employed instead of a monolithic approach.
In this scheme, the energy functional is alternatively minimized with respect to $\mathbf{u}$ and $d$.
The damage variable is subjected to an irreversibility condition $\dot{d} \geq 0$ to prevent healing.
The two resulting sets of equations read,
\begin{equation}
  \frac{\partial E_{l_0}(\mathbf{u}, d)}{\partial \mathbf{u}} = 0,
\end{equation}
and
\begin{equation}
  \frac{\partial E_{l_0}(\mathbf{u}, d)}{\partial d} \geq 0, \quad
  \frac{\partial E_{l_0}}{\partial d}(d_n-d_{n-1})=0, \quad d_n\geq d_{n-1}.
  \label{eq:e_min_d}
\end{equation}

\FloatBarrier
\subsection{Separation of tension and compression}
The original formulations in Equations~(\ref{eq:variational_griffith}) and~(\ref{eq:phase-field_regularization}) do not distinguish between the contribution of tension and compression in the energy available to propagate cracks.
As a result, cracks may appear in compressive regions, and interpenetration of crack faces is not prevented.
A modification of the elastic energy, which splits the strain energy into contributions from tension and compression, was proposed in~\cite{amor_regularized_2009}:
\begin{equation}
  E_{\text{ela.}} = \int_{\Omega} \left( g(d)\Psi^{+}(\boldsymbol{\varepsilon}) + \Psi^{-}(\boldsymbol{\varepsilon}) \right) d\Omega
\end{equation}

Here, the strain energy density is decomposed into contributions from shear and positive volumetric changes, and from negative volumetric changes:
\begin{equation}
  \psi^{+}(\boldsymbol{\varepsilon}) = \frac{1}{2} \kappa \langle \tr \boldsymbol{\varepsilon} \rangle_{+}^2 
  + \mu \boldsymbol{\varepsilon}^{dev}:\boldsymbol{\varepsilon}^{dev}, \qquad
  \psi^{-}(\boldsymbol{\varepsilon}) = \frac{1}{2} \kappa \langle \tr \boldsymbol{\varepsilon} \rangle_{-}^2.
\end{equation}

Other splitting strategies have been proposed, such as the spectral split~\cite{miehe_phase_2010} and the alternative decomposition presented in~\cite{vicentini_decomposition_2024}. The importance of separating tension and compression will be discussed in the outlook section of the article, but we will not use the split in the simulation results presented in this work.

\FloatBarrier
\subsection{Mass degradation}
The degradation of stiffness in Equation~(\ref{eq:phase-field_regularization}) is accompanied by a corresponding reduction of the elastic wave speed in the damaged zone:
\begin{equation}
     c_s=\sqrt{\frac{g(d)\mu}{\rho}}.
\end{equation}
An interesting approach, which does not lead to this alteration of the wave velocity, is proposed in Ref.~\cite{chen_instability_2017, VASUDEVAN2021104372}.
The authors of these works argue that fracture is not necessarily associated with a loss of elastic wave velocity in the process zone, and suggest a strategy to preserve it by scaling the mass density in the same way as the stiffness.
If the stiffness is scaled by the degradation function $g(d)$, the mass density is similarly scaled as $\rho(d) = g(d) \rho_0$.

This ensures that the elastic wave velocity in the process zone remains constant.
For example, the shear wave speed $c_s$ satisfies
\begin{equation}
  c_{pz} = \sqrt{\frac{\mu (d)}{\rho (d)}} = \sqrt{\frac{\mu_0 g(d)}{\rho_0 g(d)}} = c_s.
\end{equation}
Naturally, this procedure leads to a process called mass erosion, and does not satisfy the general principle of mass conservation.

The complete energy functional with this addition would be
\begin{equation}
	E_{l_0}(\disp, d) = \int_{\Omega} g(d)\Psi (\eps) \, dV
	+ \frac{G_c}{c_w}\int_{\Omega} \left( \frac{w(d)}{l_0} + l_0 \|\nabla d\|^2 \right) \, dV
    - \int_{\Omega} \frac{1}{2} g(d) \rho \dot{\disp} \cdot \dot{\disp}\, dV
    - W_{\text{ext}}(\disp),
  \label{eq:phase-field_dynamic_mass}
\end{equation}
whose derivative with respect to $d$ would give a contribution of the kinetic energy.
In this work, however, we do not include this contribution in the damage problem, as it would be if the variational form were respected, and the degradation of mass acts only on the dynamics problem.
However, this simplified form is sufficient for the purpose of this paper.

\FloatBarrier
\subsection{Numerical implementation}

The problem is discretized using the finite element method and is implemented in the open-source FEM library Akantu~\cite{Richart2024}.
The displacement $\mathbf{u}$, velocity $\dot{\mathbf{u}}$ and acceleration $\ddot{\mathbf{u}}$, satisfying the damage-dependent elastodynamic equilibrium 
\begin{equation}
  \nabla \cdot \boldsymbol{\sigma}(d) + \mathbf{b} = \rho \ddot{\mathbf{u}},
\end{equation}
with appropriate boundary conditions and the body forces $\mathbf{b}$, are updated using an explicit Newmark scheme.

The damage problem is solved once per time step using the current displacement field.
The phase-field equation is obtained from the minimization of the energy functional with respect to the damage variable $d$, Equation~(\ref{eq:e_min_d}).
Its weak form reads: \\
    Find $d \in \mathcal{D}$ such that
\begin{equation}
	\int_{\Omega} \left(-2(1-d)\Psi (\eps) v
	+ \frac{3G_c}{8} \left( \frac{1}{l_0} v + 2l_0 \nabla d \cdot \nabla v \right)\right) \, dV = 0.
  \quad \forall v \in \mathcal{D}
\end{equation}

The subsequent finite element discretization leads to the following element-level contributions to the Hessian matrix $H$ and the residual vector $b$:
\begin{align}
  H^e &= \int_{\Omega^e} \left( 2\Psi (B_u \mathbf{u}) N_d^T N_d
    + \frac{3G_c}{8} l_0 B_d^T B_d \right) \, dV, \\
    b^e &= \int_{\Omega^e} \left( 2\Psi (B_u \mathbf{u}) \frac{3G_c}{8 l_0} \right) N_d \, dV.
\end{align}
The terms $N_d$ denote the shape function matrix for the damage field, while $B_d$ and $B_u$ represent the strain-displacement matrices for the damage and displacement fields, respectively.

Following~\cite{li_gradient_2016}, the damage problem can be reformulated as a constrained quadratic optimization problem:
\begin{equation*}
  \min_{d} \quad \frac{1}{2} d^T H d - d^T b
  \quad \text{s.t.} \quad 1 \geq d \geq d_{n-1}.
\end{equation*}

This problem is solved using the Gradient Projection Conjugate Gradient (GPCG) method, as implemented in the Toolkit for Advanced Optimization (TAO) of the PETSc library~\cite{petsc-user-ref, petsc-efficient, tao-user-ref, gpcg_more_toraldo_1991}.

\FloatBarrier
\section{Dynamic fragmentation and damage spreading}
\label{sec:fragmentation_full}

\begin{figure}[ht]
  \centering
  \includegraphics{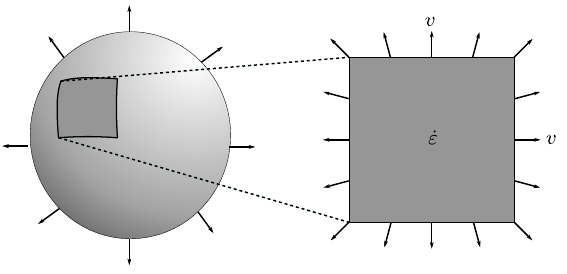}
  \caption{Geometry and boundary conditions for the fragmentation of an expanding spherical membrane, and the equivalent 2d problem, a plate in radial expansion.}
  \label{fig:fragmentation_geo}
\end{figure}

As an initial demonstration of dynamic fragmentation, and to illustrate the difficulties that arise when using the classical phase-field formulation without mass degradation, we consider the case of an expanding spherical membrane, pictured in Figure~\ref{fig:fragmentation_geo}.
A square portion of the sphere, of size $\qtyproduct{1 x 1}{\centi\meter}$, is extracted and approximated in two dimensions as a square plate in plane-strain condition and subjected to radial loading.
The geometry is discretized using 195332 linear triangular elements.

A velocity gradient is prescribed throughout the domain to impose a uniform initial strain rate $\dot{\varepsilon}$ at the beginning of the simulation.
This loading is then maintained by incrementally adjusting the imposed displacement at the boundary to preserve the target strain rate during the dynamic evolution.

The material parameters are: a Young’s modulus $E = \SI{370}{\giga\pascal}$, a Poisson's ratio $\nu = 0.22$, and a density $\rho = \SI{3900}{\kilogram\per\meter\cubed}$.
The regularization length is set to $l_0 = \SI{0.111}{\milli\meter}$.
 The critical energy release rate $G_c$ is modelled via a Weibull distribution with probability density function
\begin{equation}
f(x;\lambda,k) =
  \begin{cases}
    \frac{k}{\lambda} \left( \frac{x}{\lambda} \right)^{k-1}e^{-(x/\lambda)^k}, & x \geq 0, \\
    0, & x < 0,
  \end{cases}
\end{equation}
with the scale parameter $\lambda = 50$ and the shape parameter $k = 10$ to account for material heterogeneity.
These parameters yield a mean $\bar{G_c}=\SI{47.6}{\joule\per\square\meter}$ and a standard deviation of $\SI{5.7}{\joule\per\square\meter}$.

As explained in the previous section, the problem is solved using an explicit Newmark scheme with a time step $\Delta t = \SI{3.367d-3}{\nano\second}$ and the damage problem being solved at each time step.

\begin{figure}[ht]
    \centering
  \begin{subfigure}[b]{0.23\linewidth}
    \centering
    \includegraphics[width=\linewidth]{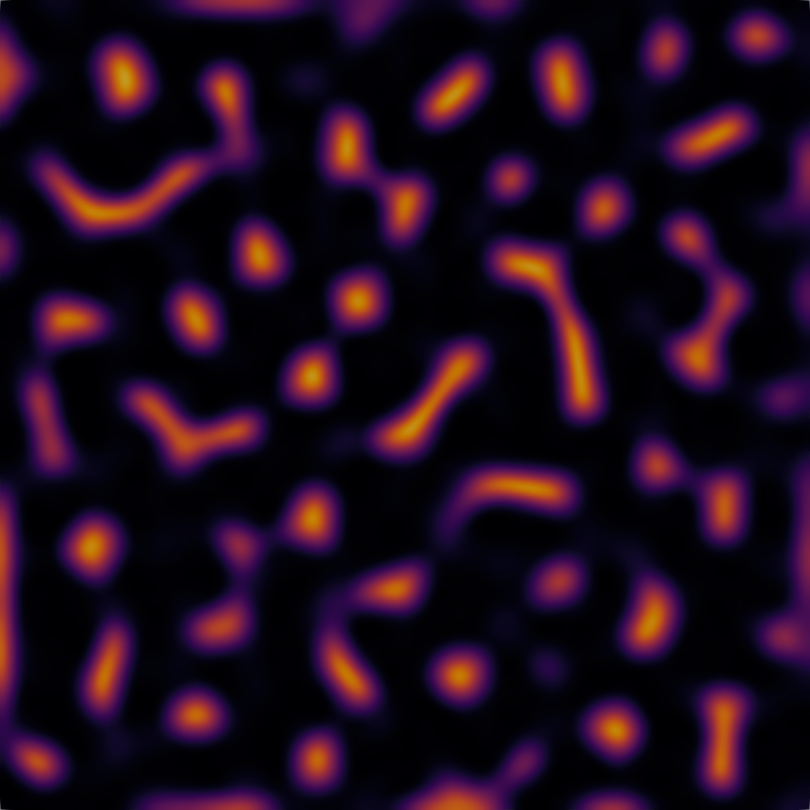}
    \caption{$t = \SI{56.6}{\nano\second}$}
  \label{fig:ex_frag_full_0}
  \end{subfigure}
  \begin{subfigure}[b]{0.23\linewidth}
    \centering
    \includegraphics[width=\linewidth]{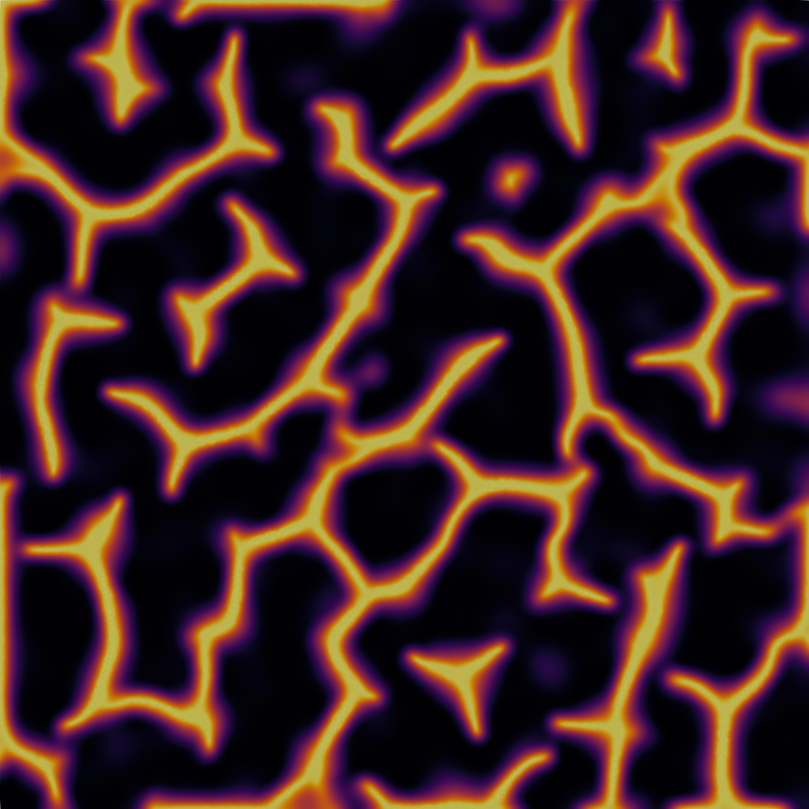}
    \caption{$t = \SI{58.9}{\nano\second}$}
  \label{fig:ex_frag_full_1}
  \end{subfigure}
  \begin{subfigure}[b]{0.23\linewidth}
    \centering
    \includegraphics[width=\linewidth]{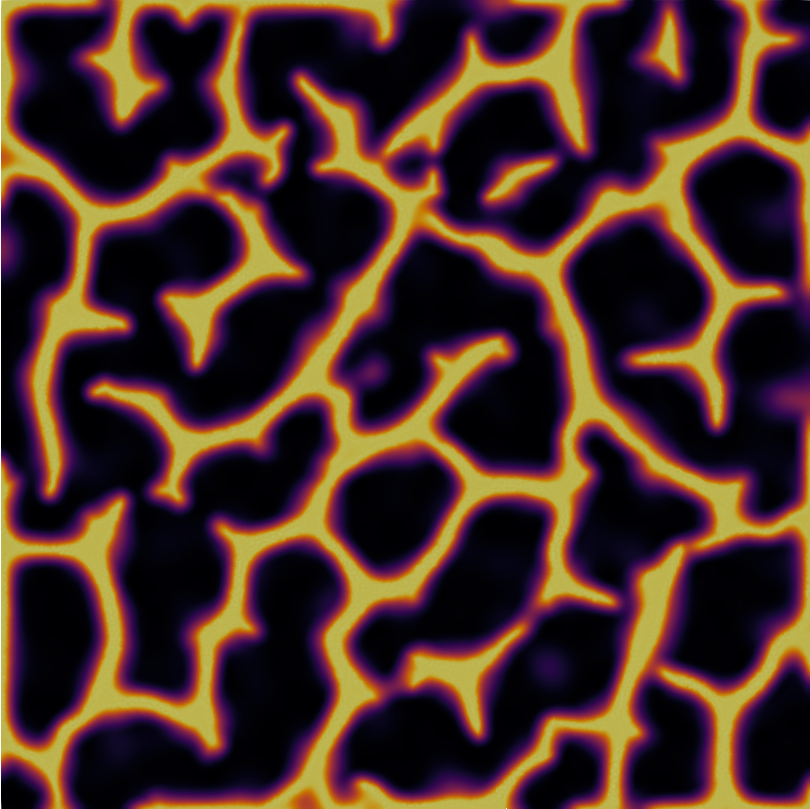}
    \caption{$t = \SI{62.6}{\nano\second}$}
  \label{fig:ex_frag_full_2}
  \end{subfigure}
  \begin{subfigure}[b]{0.23\linewidth}
    \centering
    \includegraphics[width=\linewidth]{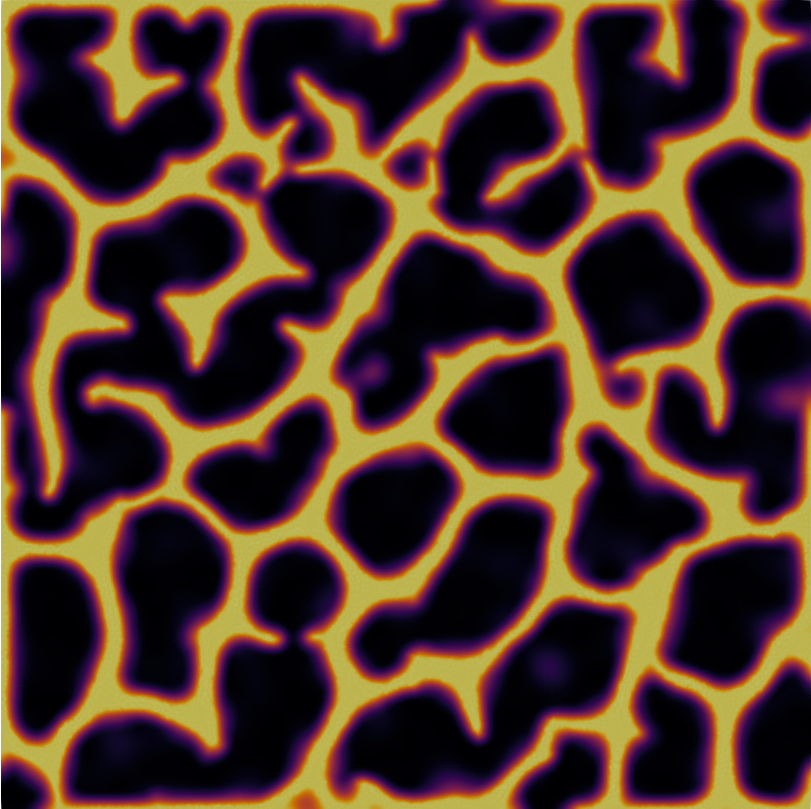}
    \caption{$t = \SI{70.7}{\nano\second}$}
  \label{fig:ex_frag_full_3}
  \end{subfigure}
  \vspace{1em}
  \begin{subfigure}[b]{0.23\linewidth}
    \centering
    \includegraphics[width=\linewidth]{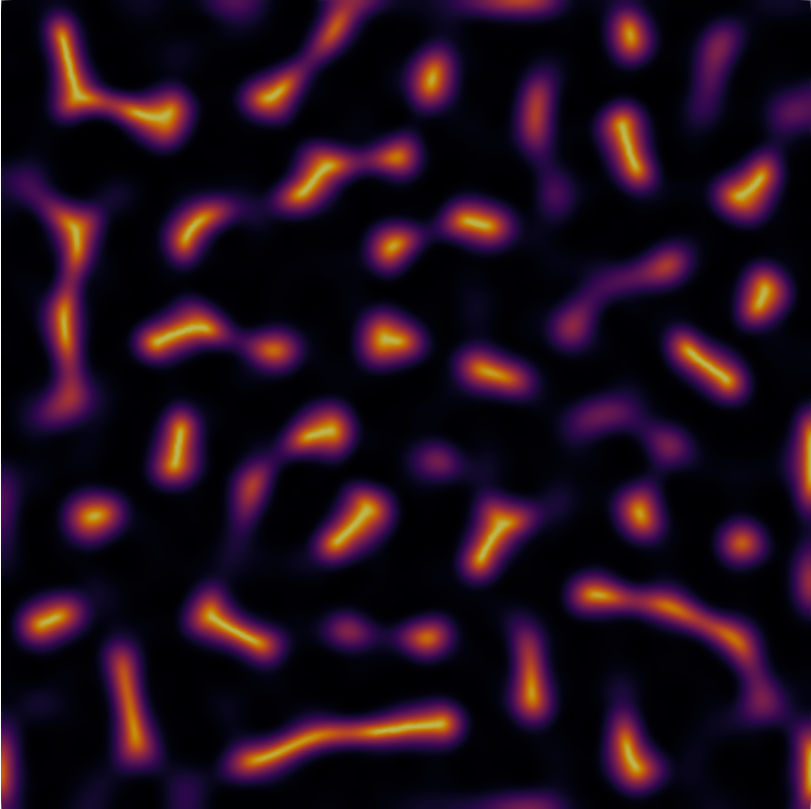}
    \caption{$t = \SI{56.6}{\nano\second}$}
  \label{fig:ex_frag_deg_0}
  \end{subfigure}
  \begin{subfigure}[b]{0.23\linewidth}
    \centering
    \includegraphics[width=\linewidth]{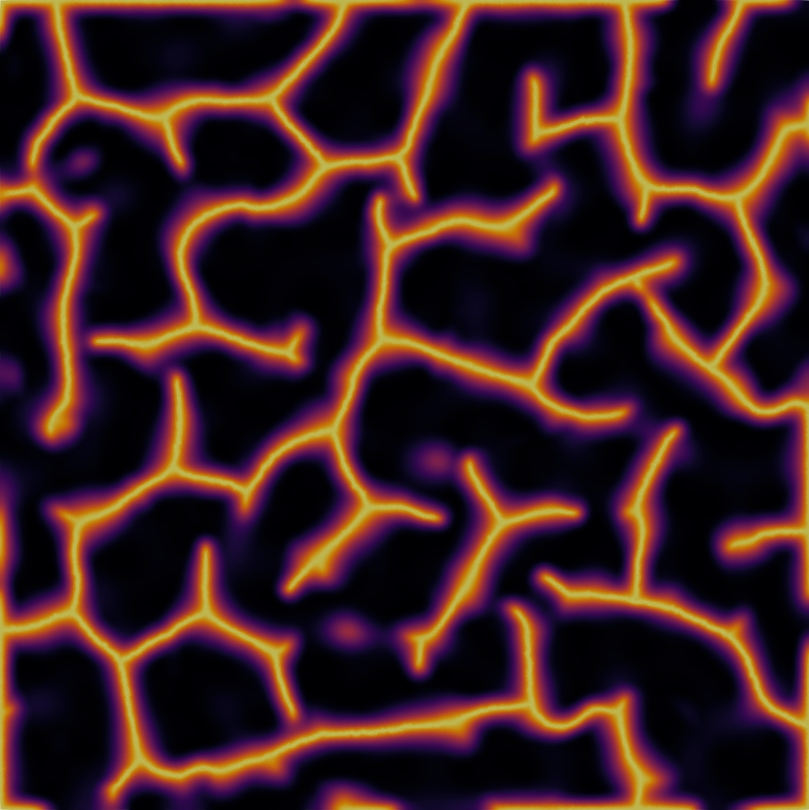}
    \caption{$t = \SI{58.9}{\nano\second}$}
  \label{fig:ex_frag_deg_1}
  \end{subfigure}
  \begin{subfigure}[b]{0.23\linewidth}
    \centering
    \includegraphics[width=\linewidth]{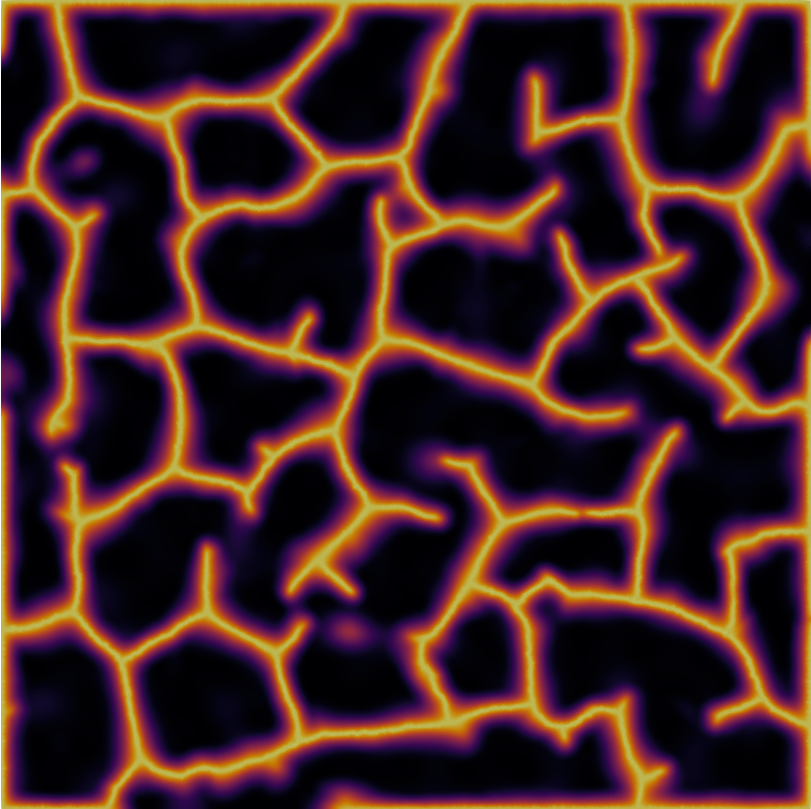}
    \caption{$t = \SI{62.6}{\nano\second}$}
  \label{fig:ex_frag_deg_2}
  \end{subfigure}
  \begin{subfigure}[b]{0.23\linewidth}
    \centering
    \includegraphics[width=\linewidth]{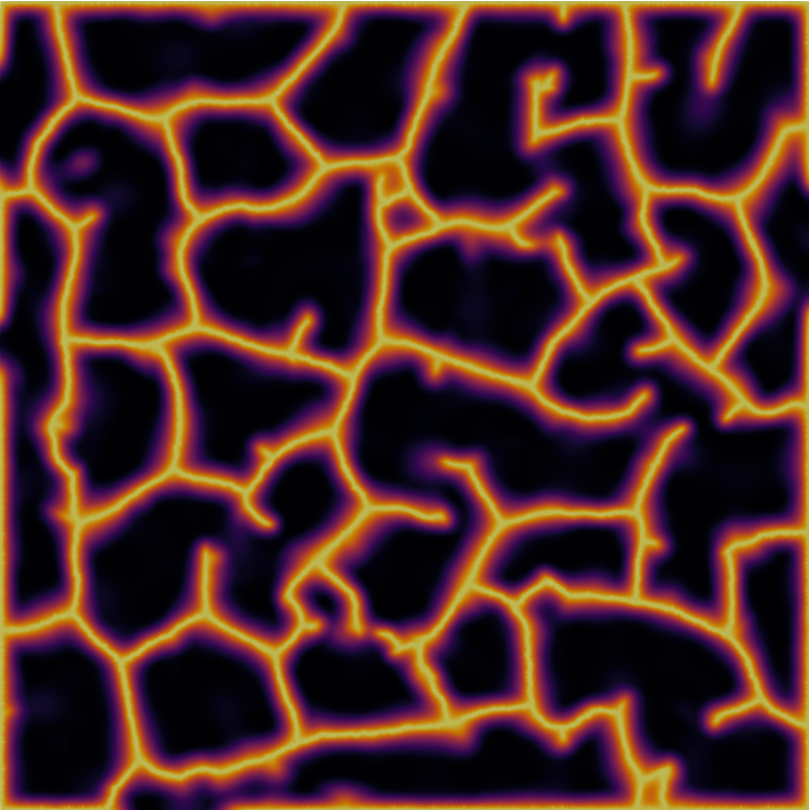}
    \caption{$t = \SI{70.7}{\nano\second}$}
  \label{fig:ex_frag_deg_3}
  \end{subfigure}
  \captionsetup{subrefformat=parens}
  \caption{\subref{fig:ex_frag_full_0}-\subref{fig:ex_frag_full_3} Damage field without mass degradation at four different time steps. \subref{fig:ex_frag_deg_0}-\subref{fig:ex_frag_deg_3} Damage field with mass degradation at four different time steps.}
  \label{fig:ex_frag}
\end{figure}

The crack pattern obtained for a strain rate of $\dot{\varepsilon}=\SI{100}{\per\second}$ and at four different times is shown in Figures~\ref{fig:ex_frag_full_0}-\subref{fig:ex_frag_full_3}.
We observe the nucleation, propagation, and coalescence of multiple fractures, ultimately leading to the formation of fragments.
However, the crack propagation is accompanied by a substantial widening of the damage bands, reaching several times their initial width.
As a result, the behaviour within the damaged zones becomes difficult to interpret, and the precise geometry of the resulting fragments cannot be reliably determined.
Furthermore, a significant fraction of material is effectively lost within these highly damaged regions, which may compromise the accuracy of the predicted fragment masses.

The next sections investigate the mechanisms responsible for this excessive increase in crack thickness and explore the use of mass degradation as a strategy to obtain qualitatively more interpretable results in dynamic fragmentation. To illustrate our purpose, the same simulation test is now repeated with mass degradation.
The resulting crack pattern is shown in Figures~\ref{fig:ex_frag_deg_0}-\subref{fig:ex_frag_deg_3}.
As observed, the widths of the damage bands are better controlled, allowing the shapes of the resulting fragments to be identified with greater accuracy.

\FloatBarrier
\section{Interaction of elastic waves with damage}
\label{sec:1d_bar}
\subsection{Wave reflection at a damaged free boundary}
\label{sec:bar_reflect}

\begin{figure}[ht]
  \centering
  \includegraphics{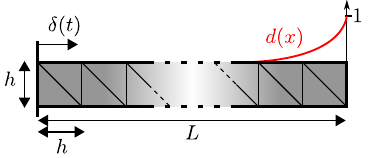}
  \caption{Geometry and boundary condition of the 1D bar test. An impulse is sent from the left toward a free boundary (right), with or without damage.}
  \label{fig:bar_reflection}
\end{figure}

The objective of this section is to compare the behaviour of a free boundary with that of a damaged boundary under dynamic loading.
To investigate the reason for the observed widening of the damage bands, we simplify the problem to a pseudo-1D bar, illustrated in Figure~\ref{fig:bar_reflection}.

The bar, of length $L=\SI{5}{\milli\meter}$, is discretized using linear triangular elements of base and height $h = \SI{5}{\micro\meter}$, which is also the height of the bar.
For the material parameters, we choose a Young's modulus $E = \SI{275}{\giga\pascal}$, a Poisson's ratio $\nu = 0$, and a density of $\rho = \SI{2750}{\kilogram\per\meter\cubed}$.
The fracture related parameters are $G_c = \SI{200}{\joule\per\square\meter}$ and $l_0 = \SI{200}{\micro\meter}$, which is much larger than the mesh size and ensures a good resolution of the response within the damaged region.
The right end of the bar is left free to undergo horizontal motion.
Damage is prescribed to be equal to~1 at the right boundary, and the corresponding optimal damage profile is obtained by solving the damage problem.
A sinusoidal impulse $\delta(t) = 0.08 \frac{E}{\rho}\Delta t \sin{\frac{\pi t}{1200 \Delta t}}$ is applied at the left end of the pseudo-1D bar for the 600 first time steps, as shown in Figure~\ref{fig:bar_reflection}.
The amplitude of the impulse is chosen sufficiently small to prevent the nucleation of additional damage.
The subsequent dynamic simulation is carried out using the same explicit time-integration scheme, with a time step $\Delta t = \SI{96.65}{\nano\second}$ for the solid model, while the damage field is still updated at each step to ensure that no damage develops in the intact part.

\begin{figure}[ht]
	\centering
	\begin{subfigure}{0.48\linewidth}
		\centering
		\includegraphics{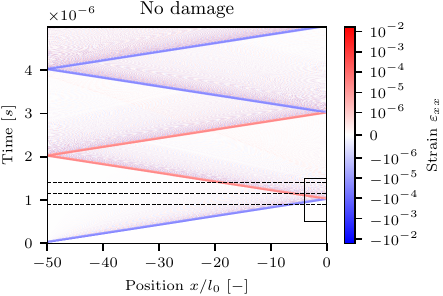}
        \caption{}
        \label{fig:strain_reflect_ref_large}
	\end{subfigure}
	\begin{subfigure}{0.48\linewidth}
		\centering
		\includegraphics{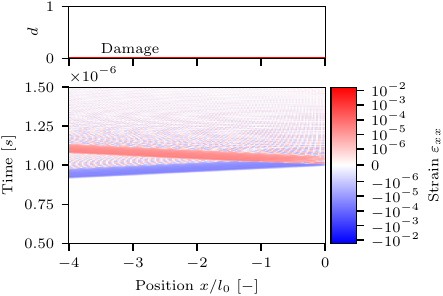}
        \caption{}
        \label{fig:strain_reflect_ref_zoom}
	\end{subfigure}
    \captionsetup{subrefformat=parens}
  \caption{\subref{fig:strain_reflect_ref_large} c-t diagram of the reference case without damage and \subref{fig:strain_reflect_ref_zoom} close up on the highlighted rectangular area. We observe a sharp reflection of the wave at the free boundary with a change of sign.}
  \label{fig:strain_reflect_ref}
\end{figure}

\begin{figure}[ht]
  \centering
  \includegraphics{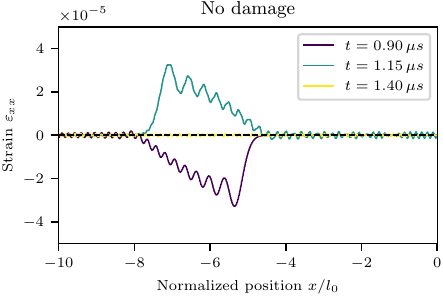}
  \caption{Strain as a function of the position close to the free boundary for the reference case without damage at three different times marked by dashed lines in Figure~\ref{fig:strain_reflect_ref_large}.}
  \label{fig:strain_reflect_ref_wave}
\end{figure}

A reference simulation is first conducted without damage to establish a baseline for the subsequent analysis.
The results are presented in Figure~\ref{fig:strain_reflect_ref} as a c–t diagram showing the strain as a function of time and of the position along the bar, normalized by the regularization length.
The compressive wave propagates from the left boundary toward the right.
As expected, upon reaching the free end, the incident compressive wave is sharply reflected as a tensile wave, producing a clear change of sign in the strain field.
A small amount of numerical noise (dispersion) is visible after the passage of the impulse, but it does not affect the overall behaviour of the solution.
This can also be seen in Figure~\ref{fig:strain_reflect_ref_wave}, which shows the strain as a function of position at three different times.
The reflected wave maintains its shape and amplitude, and the noise remains of small amplitude after the main wave passes.

\FloatBarrier

\begin{figure}[ht]
		\centering
	\begin{subfigure}{0.48\linewidth}
		\centering
		\includegraphics{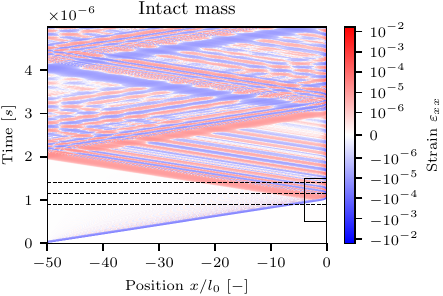}
        \caption{}
        \label{fig:strain_reflect_full_large}
	\end{subfigure}
	\begin{subfigure}{0.48\linewidth}
		\centering
		\includegraphics{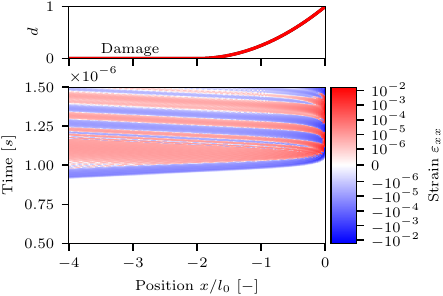}
        \caption{}
        \label{fig:strain_reflect_full_zoom}
	\end{subfigure}
    \captionsetup{subrefformat=parens}
  \caption{\subref{fig:strain_reflect_full_large} c-t diagram with damage but without mass degradation and \subref{fig:strain_reflect_full_zoom} magnification on one reflection at the damaged end with the corresponding damage value. 
    The reflection is not sharp any more and parts of the waves are trapped in the damaged area.}
  \label{fig:strain_reflect_full}
\end{figure}

The simulation is then repeated with a damaged boundary, first without mass degradation, as shown in Figure~\ref{fig:strain_reflect_full}.
The same impulse propagates from the left toward the right.
Upon reaching the damaged end, the wave is progressively slowed down and the strain amplitude increases by several orders of magnitude.
A portion of the wave is reflected earlier than in the reference case, indicating a modification of the effective boundary response.
Following the arrival of the wave, the damaged region enters what may be described as a ringing regime: waves become trapped within the damaged zone, and small oscillations persist over time.
The damaged end continues to emit alternating tensile and compressive waves, which subsequently interact with the next incoming main wave.
Thus, even in the absence of additional damage spreading, the behaviour of the damaged boundary is problematic, as it does not reproduce the expected response of a free boundary.

\begin{figure}[ht]
	\centering
	\begin{subfigure}{0.48\linewidth}
		\centering
		\includegraphics{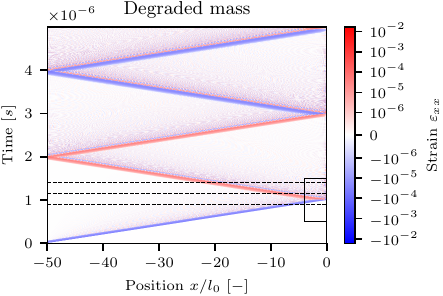}
        \caption{}
        \label{fig:strain_reflect_deg_large}
	\end{subfigure}
	\begin{subfigure}{0.48\linewidth}
		\centering
		\includegraphics{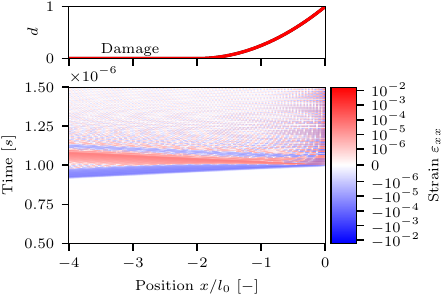}
        \caption{}
        \label{fig:strain_reflect_deg_zoom}
	\end{subfigure}
    \captionsetup{subrefformat=parens}
  \caption{\subref{fig:strain_reflect_deg_large} c-t diagram with damage and mass degradation and \subref{fig:strain_reflect_deg_zoom} magnification on one reflection at the damaged end.
    A sharp reflection is recovered, closer to the reference case.
    Note that the reflection is not perfect, as a part of the reflected wave does not change sign.
  }
  \label{fig:strain_reflect_deg}
\end{figure}

\begin{figure}[ht]
	\begin{subfigure}{0.48\linewidth}
		\centering
		\includegraphics{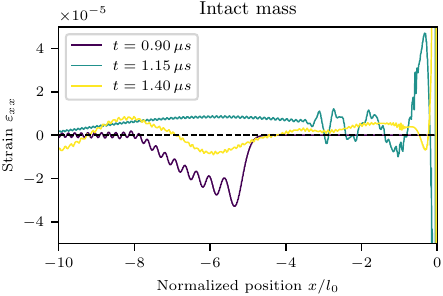}
        \caption{}
    \label{fig:strain_reflect_full_wave}
	\end{subfigure}
	\begin{subfigure}{0.48\linewidth}
		\centering
		\includegraphics{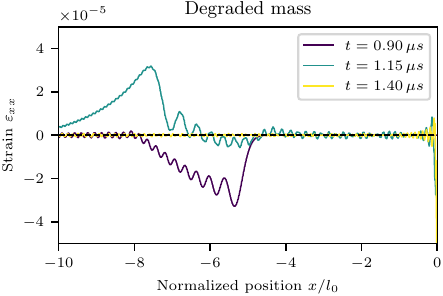}
        \caption{}
    \label{fig:strain_reflect_deg_wave}
	\end{subfigure}
    \captionsetup{subrefformat=parens}
  \caption{Strain as a function of the position close to the damaged boundary for the three different times marked by dashed lines in Figures~\ref{fig:strain_reflect_full_large} and~\ref{fig:strain_reflect_deg_large}, \subref{fig:strain_reflect_full_wave} without mass degradation and \subref{fig:strain_reflect_deg_wave} with mass degradation.}
  \label{fig:strain_reflect_wave}
\end{figure}

The simulation is again repeated with mass degradation, as shown in Figure~\ref{fig:strain_reflect_deg}.
In this case, a sharp reflection is recovered, much closer to the reference free-boundary response.
The wave is no longer slowed down, as the wave velocity is preserved within the damaged zone, and little to no increase in strain amplitude is observed.
Figure~\ref{fig:strain_reflect_wave} shows the strain as a function of position at three different times, respectively without and with mass degradation.
The deformation of the reflected wave can clearly be seen in the first case, along with the presence of residual oscillations after the main wave passage.
The right part of the figure shows a shape of the reflected wave much closer to the original impulse.
The behaviour with loss of mass is not entirely ideal, however: a small portion of the reflected wave does not undergo a complete sign reversal, and the reflection still occurs slightly earlier than expected.
Nevertheless, these results demonstrate that preserving the wave speed within the damaged region is essential for damage to accurately approximate a free boundary, thereby providing a more physically consistent representation of a fracture.

\FloatBarrier
\subsection{Damage spreading mechanism}
\label{sec:bar_oscillation}

We now investigate the damage spreading mechanism by considering the oscillations of a bar released from an initially compressed state.
The configuration is analogous to the one introduced in Section~\ref{sec:bar_reflect}, except that we now consider twice the length of the original bar.
The material parameters are identical to those used in the previous section, except for the regularization length, which is decreased to $l_0 = \SI{20}{\micro\meter}$, a value still larger than the mesh size.
The bar is initially compressed from both ends by imposing a displacement $\delta = \SI{15}{\micro\meter}$.

After applying the imposed displacement, we first solve the problem with a single static step without damage to obtain the stress and displacement fields in the bar.
The two ends of the bar are then released, and the system is free to evolve dynamically via the same explicit scheme with a time step $\Delta t = \SI{96.65}{\nano\second}$.
Two release waves propagate from the ends of the bar and superpose at the centre where they reach a sufficient deformation to nucleate a crack and the simulation is continued for a few oscillations.
This test is performed twice, first without and then with mass degradation.

\begin{figure}[ht]
  \centering
  \includegraphics{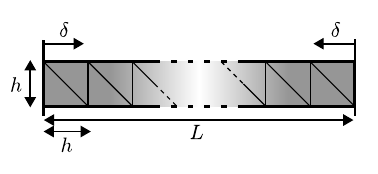}
  \caption{Geometry and boundary conditions for the oscillation test.}
  \label{fig:bar_oscillation}
\end{figure}

\begin{figure}[ht]
	\centering
	\begin{subfigure}{0.49\linewidth}
		\centering
		\includegraphics{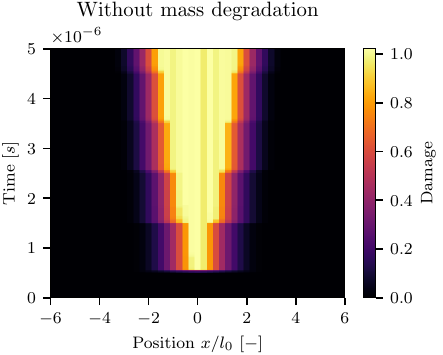}
        \caption{}
        \label{fig:oscillation_ct_full}
	\end{subfigure}
	\begin{subfigure}{0.49\linewidth}
		\centering
		\includegraphics{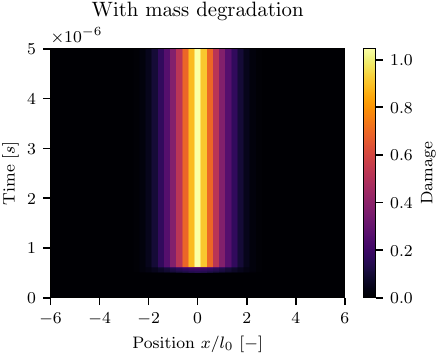}
        \caption{}
        \label{fig:oscillation_ct_deg}
	\end{subfigure}
    \captionsetup{subrefformat=parens}
  \caption{c-t diagram of the damage as a function of time and position around the centre of the bar. \subref{fig:oscillation_ct_full} Without mass degradation an increase in damage with each incoming wave is observed, while \subref{fig:oscillation_ct_deg} with mass degradation a stable damage profile is maintained.}
  \label{fig:oscillation_ct_damage}
\end{figure}

Figure~\ref{fig:oscillation_ct_damage} presents the results in the form of a c–t diagram.
The heatmap displays the damage field as a function of time and position, normalized by the regularization length $l_0$.
The view is restricted to the central region of the bar, where wave superposition occurs, and damage nucleation takes place.
In both simulations, damage initiates when the first waves reach the centre, with an initial width close to the theoretical value of $4 l_0$.
In the case without mass degradation (left), each subsequent oscillation leads to a progressive increase in damage.
In contrast, the case with mass degradation (right) exhibits a stable damage profile throughout the simulation.

\begin{figure}[ht]
	\centering
	\begin{subfigure}{0.49\linewidth}
		\centering
		\includegraphics{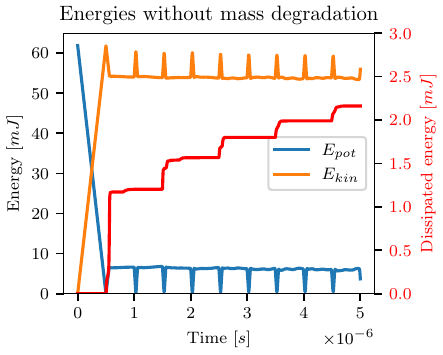}
        \caption{}
        \label{fig:oscillation_energy_full}
	\end{subfigure}
	\begin{subfigure}{0.49\linewidth}
		\centering
        \includegraphics{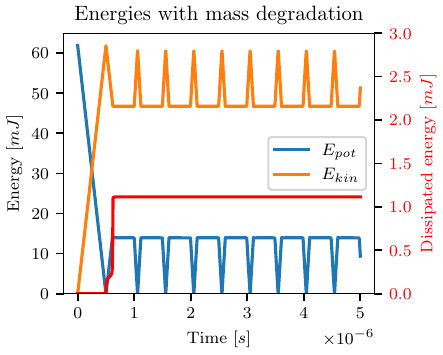}
        \caption{}
        \label{fig:oscillation_energy_deg}
	\end{subfigure}
    \captionsetup{subrefformat=parens}
    \caption{Evolution of the potential, kinetic and dissipated energy as a function of time for the oscillation test. \subref{fig:oscillation_energy_full} The plot for the simulation without mass degradation shows the increase in dissipated energy corresponding to the widening of the damage band.
    \subref{fig:oscillation_energy_deg} With mass degradation, the dissipated energy reaches a steady state.}
    \label{fig:oscillation_energy}
\end{figure}

This behaviour is further illustrated in Figure~\ref{fig:oscillation_energy}, which shows the evolution of the potential, kinetic, and dissipated energy over time.
Both simulations begin with a conversion of potential to kinetic energy as the bar is released.
Upon arrival of the waves at the centre of the bar, damage nucleates and the dissipated energy increases.
This event coincides with a reduction in kinetic energy and a corresponding rise in potential energy due to wave reflection.
The kinetic energy remains higher in the simulation without mass degradation.
This can be attributed to a cleaner reflection of the wave in the simulation with mass degradation, in contrast to the perturbations present in the damaged zone in the first case.
As expected, the simulation without mass degradation exhibits an incremental increase in dissipated energy with each oscillation.

\begin{figure}[ht]
    \centering
  \begin{subfigure}{0.49\linewidth}
    \centering
    \includegraphics{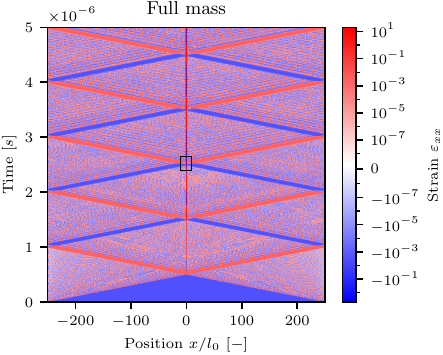}
    \caption{}
    \label{fig:strain_oscillation_full}
  \end{subfigure}
  \begin{subfigure}{0.49\linewidth}
    \centering
    \includegraphics{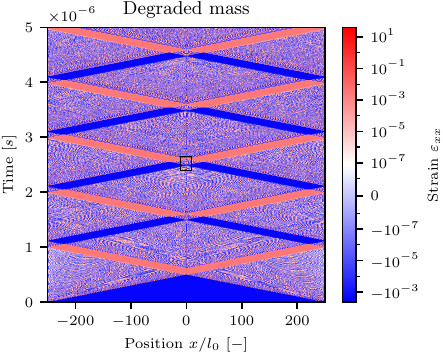}
    \caption{}
    \label{fig:strain_oscillation_deg}
  \end{subfigure}
  \captionsetup{subrefformat=parens}
  \caption{c-t diagram of the strain in the whole bar length. \subref{fig:strain_oscillation_full} The case without mass degradation shows a shorter wave span than~\subref{fig:strain_oscillation_deg} the case with mass degradation. A high strain area appears around the centre, corresponding to the damaged zone. This area increases over time in~\subref{fig:strain_oscillation_full}. Note the difference in scale for the compressive values. The magnification windows in black are shown in the next figure.}
  \label{fig:strain_oscillation}
\end{figure}

Figure~\ref{fig:strain_oscillation} presents the c–t diagram of the strain along the entire length of the bar for both simulations.
In both cases, the initially compressed bar relaxes as the release waves propagate from the ends toward the centre.
Subsequently, the bar oscillates, with alternating tensile and compressive waves reflecting at the centre and at the boundaries.

In the simulation without mass degradation, the strain values in the middle of the bar reach higher amplitudes (with an absolute maximum for $\varepsilon_{xx} = 0.9$ versus $\varepsilon_{xx} = 0.29$ with mass degradation), and more damage develops during the early stages.
As a result, more energy is trapped in the damaged zone and less of the waves are effectively reflected, leading to a thinner spatial span of the wave patterns.
Moreover, the region of high strain grows over time when mass degradation is not applied.

\begin{figure}[ht]
  \begin{subfigure}{0.49\linewidth}
    \centering
    \includegraphics{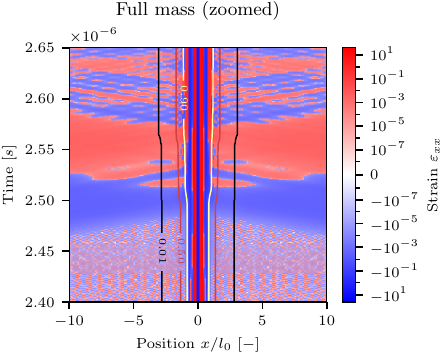}
    \caption{}
    \label{fig:strain_oscillation_full_zoom}
  \end{subfigure}
  \begin{subfigure}{0.49\linewidth}
    \centering
    \includegraphics{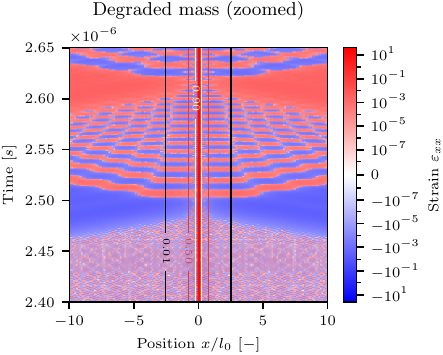}
    \caption{}
    \label{fig:strain_oscillation_deg_zoom}
  \end{subfigure}
  \captionsetup{subrefformat=parens}
  \caption{Detail of the c-t diagram of the strain around the centre of the bar for one pair of reflected waves. The contour lines represent the damage levels $d=0.01$, $d=0.5$ and $d=0.9$. \subref{fig:strain_oscillation_full_zoom} Without mass degradation, the value of the strain reach a high value associated with high damage. This zone of high strain grows as the waves are reflected. \subref{fig:strain_oscillation_deg_zoom} With mass degradation, the zone of high strain value is also present but remains contained.}
  \label{fig:strain_oscillation_zoom}
\end{figure}

A close up on a pair of reflected waves is shown in Figure~\ref{fig:strain_oscillation_zoom}.
With mass degradation, the transition from compression to tension remains smooth as the waves propagate through the damaged zone.
Without mass degradation, however, the strain increases in the damaged region.
This elevated strain promotes additional damage, as evidenced by the earlier outward movement of the $d = 0.5$ isoline, pushing in turn the higher and lower damage valued zones outward.

These findings indicate that when mass is preserved while the wave speed is reduced, elastic waves interact with the damaged zone in an undesirable manner.

\FloatBarrier
\section{2D crack propagation and branching}
\label{sec:2d}
\subsection{Single crack dynamics}

In the previous section, we identified the interaction of elastic waves with damage as a source of damage sprawl and other generally unwanted behaviour.
We showed that preserving wave velocity is a key component for obtaining reliable results under high loading.
We now shift to a two-dimensional setting to examine crack propagation with mass degradation.

\begin{figure}[ht]
  \begin{subfigure}{0.5\linewidth}
   \centering
   \includegraphics{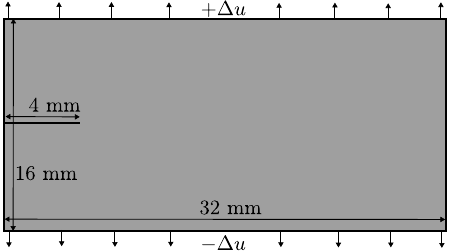}
   \caption{}
   \label{fig:notch_geo}
 \end{subfigure}
 \begin{subfigure}{0.5\linewidth}
    \centering
    \raisebox{-0.5\height}{\includegraphics[width=\linewidth]{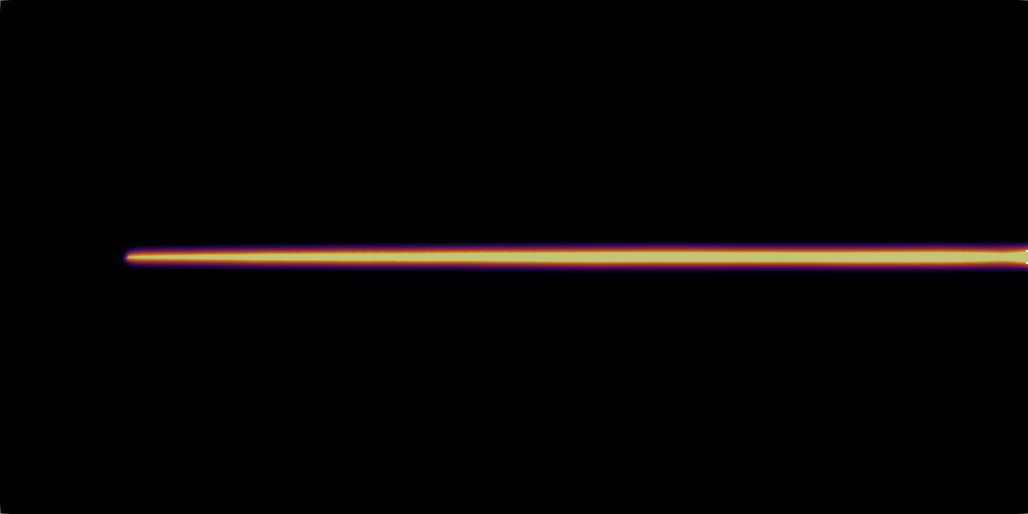}}
    \caption{}
    \label{fig:ex_single_full}
  \end{subfigure}
  \begin{subfigure}{0.5\linewidth}
    \centering
    \raisebox{-0.5\height}{\includegraphics[width=\linewidth]{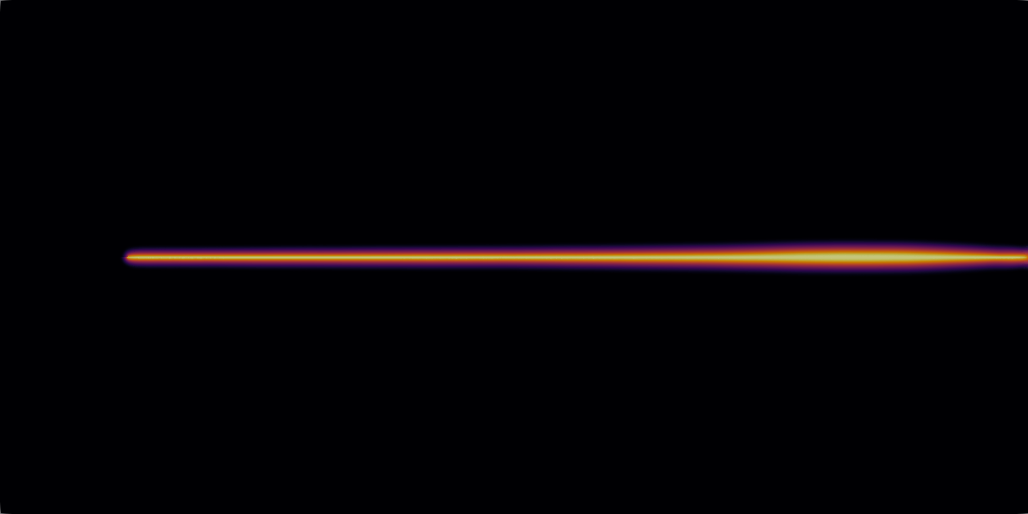}}
    \caption{}
    \label{fig:ex_single_deg}
  \end{subfigure}
  \begin{subfigure}{0.5\linewidth}
    \centering
    \raisebox{-0.5\height}{\includegraphics[width=\linewidth]{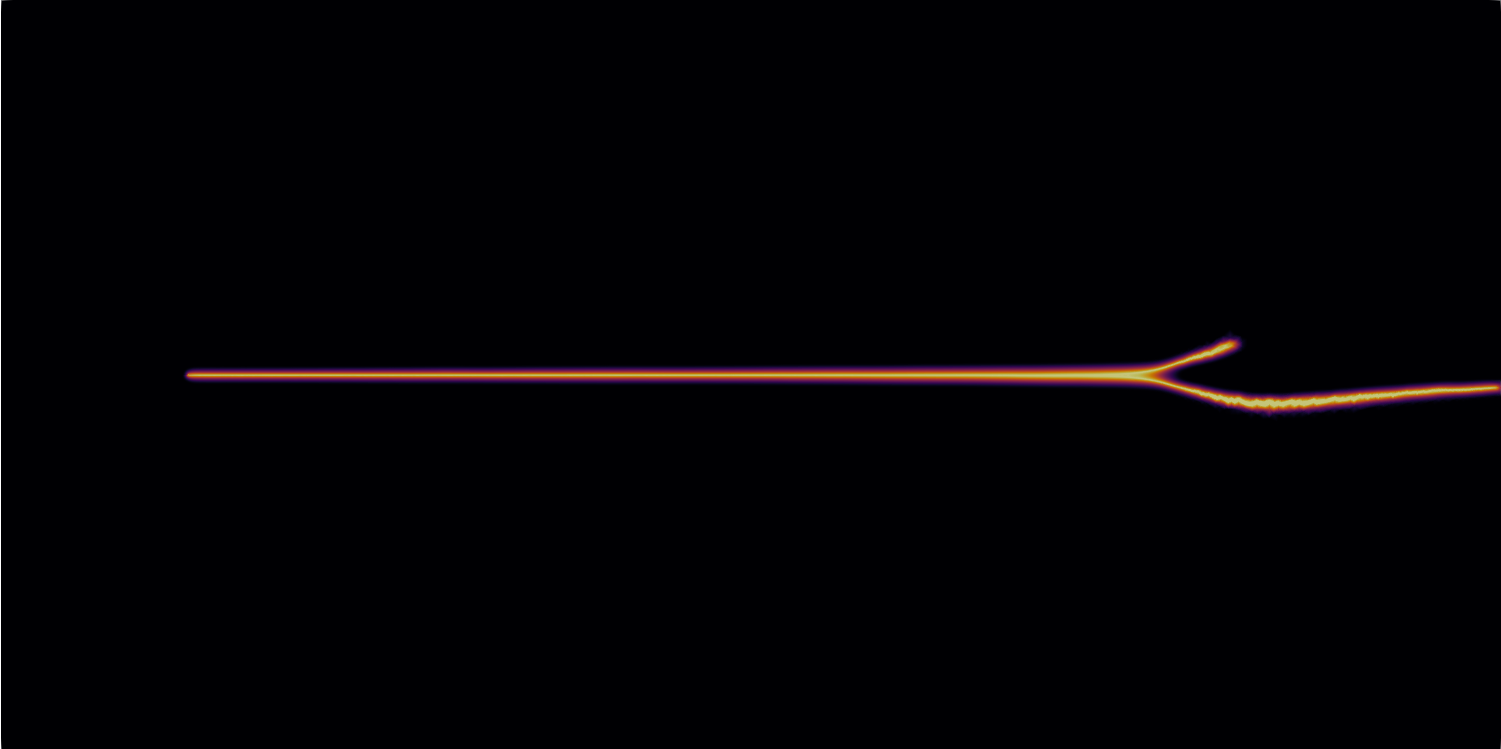}}
    \caption{}
    \label{fig:ex_single_deg_fine}
  \end{subfigure}
  \captionsetup{subrefformat=parens}
  \caption{\subref{fig:notch_geo} Geometry and boundary conditions for the crack propagation test.
    \subref{fig:ex_single_full} Example of crack propagation at $l_0 = \SI{0.2}{\milli\meter}$ and $\Delta u = \SI{36}{\micro\meter}$ without mass degradation. The crack thickness increases as the crack advances and accelerates.
    \subref{fig:ex_single_deg} Crack propagation also at $l_0 = \SI{0.2}{\milli\meter}$ and $\Delta u = \SI{36}{\micro\meter}$ with mass degradation. The crack remains thinner than without mass degradation.
    \subref{fig:ex_single_deg_fine} Crack propagation with mass erosion and a finer mesh, $l_0 = \SI{0.1}{\milli\meter}$ and $\Delta u = \SI{36}{\micro\meter}$. The crack propagates and then branches.
    }
    \label{fig:ex_single}
\end{figure}

We consider the example of a notched PMMA plate in 2D plane-stress condition, with the geometry and boundary conditions shown in Figure~\ref{fig:notch_geo}, following configurations similar to those in~\cite{bleyer_dynamic_2017, zhou_macroscopic}.
The material properties are $E = \SI{3.09}{\giga\pascal}$, $\nu = 0.35$, $G_c = \SI{300}{\joule\per\square\meter}$, and $\rho = \SI{1180}{\kilogram\per\meter\cubed}$.
The corresponding Rayleigh wave speed is $c_r=\SI{906}{\meter\per\second}$.
A prescribed displacement $\Delta u$ is imposed on the top and bottom boundaries of the plate.

As a first step, the solid mechanics problem is solved in a single static step without damage to obtain the stress and displacement fields throughout the domain.
As before, the system is then allowed to evolve dynamically, with the damage field updated once per time step based on the current displacement state.

\begin{figure}[ht]
    \centering
  \begin{subfigure}{0.49\linewidth}
    \centering
    \includegraphics{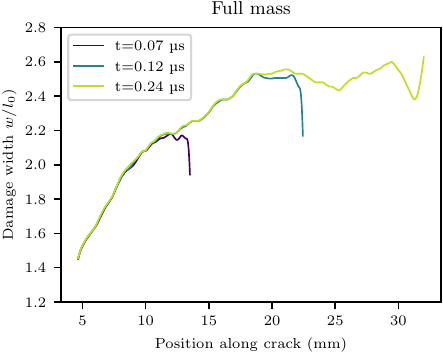}
    \caption{}
    \label{fig:damage_width_full}
  \end{subfigure}
  \begin{subfigure}{0.49\linewidth}
    \centering
    \includegraphics{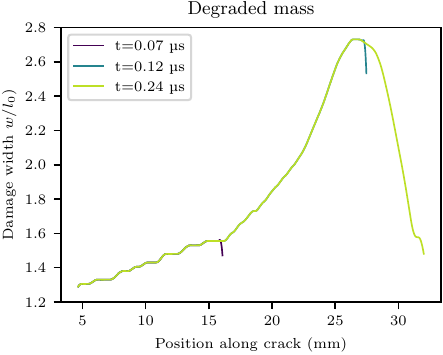}
    \caption{}
    \label{fig:damage_width_deg}
  \end{subfigure}
  \captionsetup{subrefformat=parens}
  \caption{Crack width along the crack paths in Figures~\ref{fig:ex_single_full} and~\ref{fig:ex_single_deg} for three different times, \subref{fig:damage_width_full} with and \subref{fig:damage_width_deg} without mass degradation. The curves are superposed and damage does not evolve once it is set.}
  \label{fig:damage_width_single}
\end{figure}

An example of crack propagation without mass degradation is shown in Figure~\ref{fig:ex_single_full}, for $l_0 = \SI{0.2}{\milli\meter}$ and $\Delta u = \SI{36}{\micro\meter}$.
The crack width increases rapidly as the crack accelerates, reaching values around twice its initial width—an effect already reported in~\cite{borden_phase-field_2012, li_gradient_2016}.
No crack branching is observed in this case.

The same example, but with mass degradation, is shown in Figure~\ref{fig:ex_single_deg}.
In this case, the crack thickness remains relatively low over more than half of its propagation path.
It then widens before reaching the end of the plate, accompanied by an aborted branching attempt.
Figure~\ref{fig:ex_single_deg_fine} shows that, for a regularization length reduced to $l_0 = \SI{0.1}{\milli\meter}$, branching is promoted and the wider crack near the right end in Figure~\ref{fig:ex_branch_deg} are now fully formed branches.
As will be discussed in Section~\ref{sec:2d_branching}, branching occurs more readily when mass degradation is employed.

The crack thickness as a function of the crack length at different points in time is reported in Figure~\ref{fig:damage_width_single}.
The crack width $w$ is computed in post-processing by rasterizing the damage heatmap, thresholding at a certain damage value and counting the remaining pixel in each column.
The difference in width lies primarily in the highly damaged area of the fracture, so a threshold of $d=0.5$ is chosen.
As the curves nearly superpose in time, and in opposition to the observations of the previous section, we conclude that no widening occurs after the passing of the crack. Vibrations and wave trapping are not significant in this simulation setup. Generally, the crack width is larger when there is no mass degradation.
This can be imputed to slower unloading (slower wave speeds) of the fracture faces; the additional dissipation in the width of the crack is intimately linked to a slower propagation (see crack tip position as function of time).

\begin{figure}[ht]
  \centering
  \includegraphics{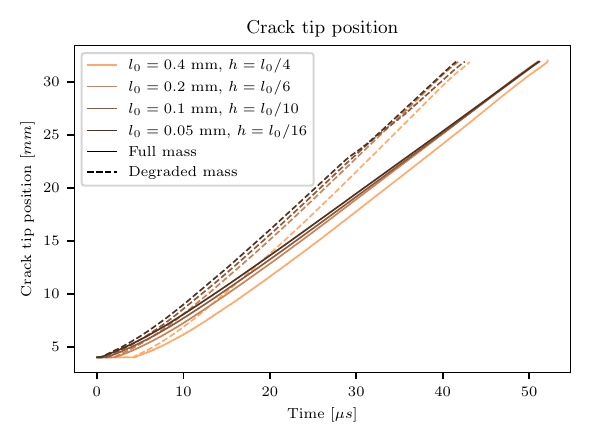}
  \caption{Crack tip position as a function of time for decreasing regularization lengths $l_0$, for the case without mass degradation (full lines) and with mass degradation (dashed lines) and for an imposed displacement $\Delta u = \SI{36}{\micro\meter}$. The slopes of the curves are much steeper in the case with mass degradation. We observe a convergence as the regularization length decreases for both cases with and without mass degradation.}
  \label{fig:tip_position}
\end{figure}

The next step is to perform a convergence analysis of crack propagation as the regularization length $l_0$ tends to zero.
The mesh element size $h$ is refined accordingly such that
\begin{equation}
l_0 \to 0 \qquad \text{and} \qquad \frac{l_0}{h} \to \infty,
\end{equation}
ensuring adequate resolution of the diffused crack as $l_0$ decreases.
From this point onward, a fixed value for $l_0$ means also a corresponding fixed mesh size $h$ whose values are displayed in the inset of Figure~\ref{fig:tip_position}.
The boundary conditions remain the same as in Figure~\ref{fig:ex_single}.
Convergence is assessed based on the crack-tip position as a function of time, the corresponding crack velocity, and the energy dissipation per unit crack extension.

The crack-tip position is defined as the rightmost point at which the damage variable satisfies $d \geq 0.9$.
The crack velocity is computed by finite differences of the crack-tip position over successive time increments.
The resulting velocity curves are subsequently smoothed using a moving-average filter.

Figure~\ref{fig:tip_position} shows the crack-tip position as a function of time for decreasing values of the regularization length $l_0$, for both the cases with and without mass degradation.
The simulations with mass degradation are shown using dashed lines.
In both sets of curves, a darker colour corresponds to a smaller value of $l_0$.
For the darkest dashed curves (smallest $l_0$), a kink appears in the trajectory.
This kink corresponds to the onset of a branching event, as seen in Figure~\ref{fig:ex_single_deg_fine}, which temporarily slows down the crack.
A short delay before crack initiation is also noticeable, and this delay decreases as $l_0$ becomes smaller; more details on this phenomenon can be found in Ref.~\cite{loiseau_initial_2025}.

\begin{figure}[ht]
    \centering
    \includegraphics{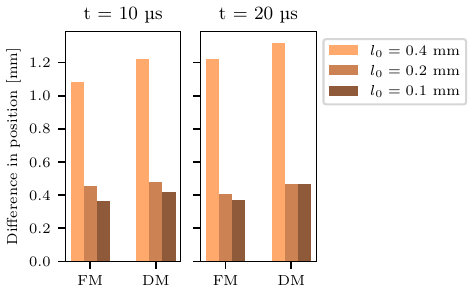}
    \caption{Difference in position from Figure~\ref{fig:tip_position} relative to the next lower $l_0$ value for two different times. FM is for full mass and DM for degraded mass. The difference in position decreases with $l_0$.}
    \label{fig:l0_convergence_bar}
\end{figure}

Remarkably, the simulations provide evidence of convergence of the phase-field approach,  as the trajectories become closer to one another with $l_0$ decreasing. To the best of our knowledge, it is the first time that convergence is reported for a dynamically propagating crack.  A second notable observation is the difference in slope between the two families of curves: the simulations with mass degradation exhibit a steeper slope, and therefore a higher crack velocity.
The convergence trend is further reported in Figure~\ref{fig:l0_convergence_bar}, where the distance between the crack tips of two consecutive decreasing regularisation lengths is presented at two different times in the simulation.

\begin{figure}[ht]
  \centering
  \includegraphics{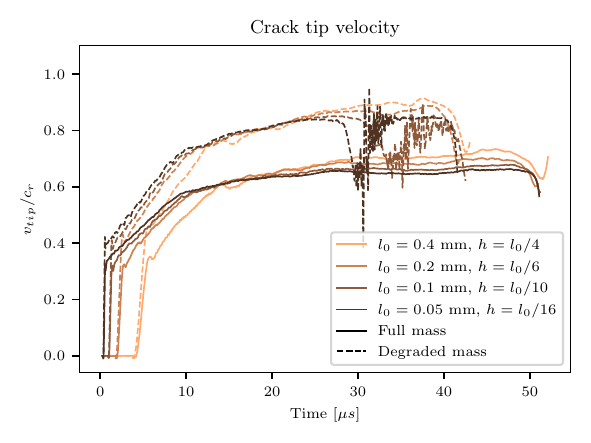}
  \caption{Crack tip velocity as a function of time, normalized by $c_r$, for $\Delta u = \SI{36}{\micro\meter}$. Without mass degradation, the dynamically propagating cracks reach a velocity of approximately $0.6 c_r$, while with mass degradation they reach a velocity of approximately $0.8 c_r$.}
  \label{fig:tip_velocity}
\end{figure}

The corresponding crack velocities are reported in Figure~\ref{fig:tip_velocity}.
Without mass degradation, the curves clearly converge to a steady velocity of approximately $0.6,c_r$.
When mass degradation is included, the fracture reaches higher velocities, exceeding $0.8 c_r$, because the elastic waves are not slowed down in the process zone and less energy is dissipated within the crack width.
The kink observed in Figure~\ref{fig:tip_position} becomes even more pronounced here, as the velocity drops when the fracture begins to branch.
This branching event also appears to occur earlier as $l_0$ decreases.

\begin{figure}[ht]
  \centering
  \includegraphics{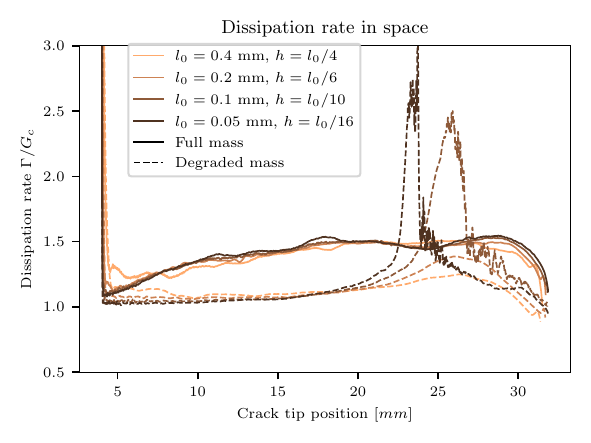}
  \caption{Dissipation rate as a function of crack length, normalized by the critical energy release rate $G_c$, and for $\Delta u = \SI{36}{\micro\meter}$. The initial large value is an artifact at the onset of crack propagation, since the damage width increases without any crack advance. When the crack propagates, and in the case without mass degradation, the dissipation rate is higher than $G_c$. Degrading the mass leads to a dissipation rate converging to $G_c$. For a fine mesh, branching can be seen as the dissipation rate jumps to values larger than twice $G_c$.}
  \label{fig:dissipation_rate_space}
\end{figure}

The dissipated energy is now examined in detail.
The total dissipated energy is computed from the damage field using the Ambrosio–Tortorelli regularization,
\begin{equation}
	E_d(d) = \frac{G_c}{c_w}\int_{\Omega} \left( \frac{w(d)}{l_0} + l_0 \|\nabla d\|^2 \right) \, dV,
  \label{eq:dissipated_energy}
\end{equation}
and the dissipation rate is defined as the increment of dissipated energy per increment of crack length,
\begin{equation}
  \Gamma = \frac{dE_d}{da}
  \label{eq:dissipation_rate}
\end{equation}
with $a$ denoting the theoretical crack length.
Note that this method of computing is not accurate after branching, as the real crack length does not correspond to the computed value.
The results are shown in Figure~\ref{fig:dissipation_rate_space}.

The very large dissipation rate observed at the beginning of the simulation is an artifact due to the initial increase of damage near the notch tip, and hence dissipation, prior to the onset of crack propagation.
Without mass degradation, once this initial transient has passed, the dissipation rate rises from $1 \times G_c$ to slightly below $1.5 \times G_c$.
The lower crack velocity in this case leads to enhanced dissipation within the crack width.

With mass degradation, the dissipation rate converges toward $G_c$ until the point at which branching initiates, at which point the dissipation goes above $2 \times G_c$ for the two lowest regularization length.
Again here, the lower rise in dissipation rate toward the end of the plate for the two lighter curves correspond to the widening of the crack observed in Figure~\ref{fig:ex_single_deg}, while the sharper peaks in the two darker curves correspond to the branching in Figure~\ref{fig:ex_single_deg_fine}.
The reduced dissipation rate favours branching, as the system seeks additional mechanisms to dissipate the required energy.
The higher crack velocity in the degraded-mass case also makes the crack more susceptible to instabilities, either numerically in this case, or due to small material heterogeneities if they were included.
Currently, it is unclear whether there can be numerical mesh convergence of dynamic instabilities such as branching events.

\FloatBarrier
\subsection{Branching crack dynamics}
\label{sec:2d_branching}

\begin{figure}[ht]
  \centering
  \begin{subfigure}{0.495\linewidth}
    \centering
    \includegraphics[width=\linewidth]{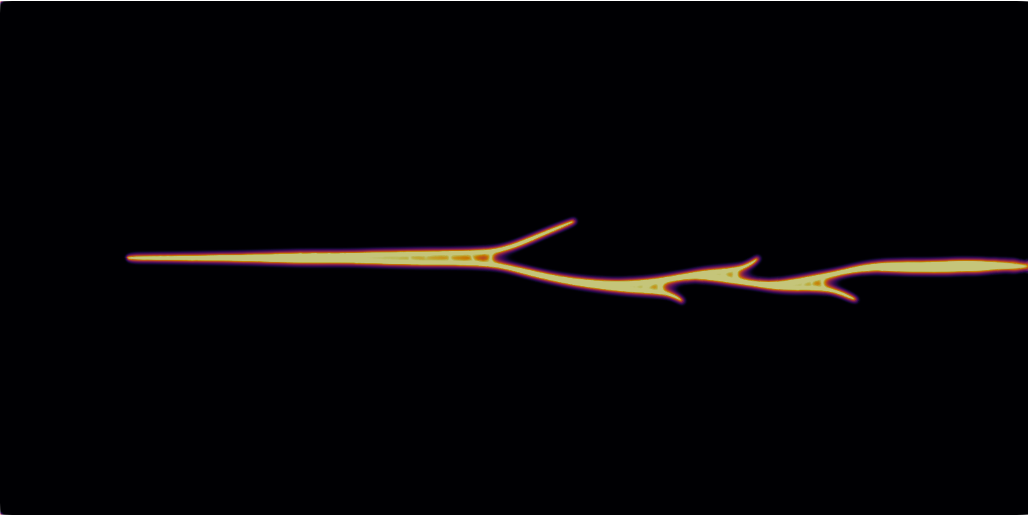}
    \caption{}
    \label{fig:ex_branch_full}
  \end{subfigure}
  \begin{subfigure}{0.495\linewidth}
    \centering
    \includegraphics[width=\linewidth]{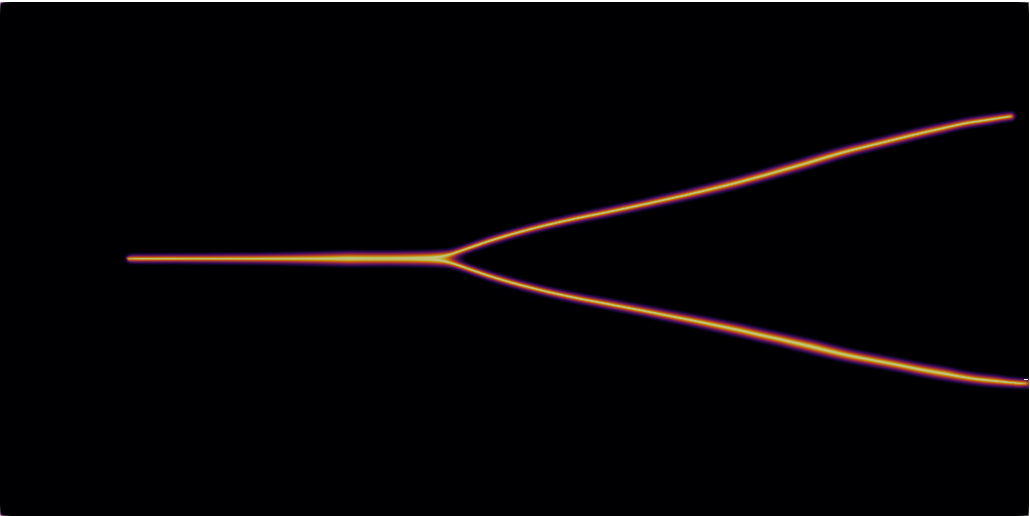}
    \caption{}
    \label{fig:ex_branch_deg}
  \end{subfigure}
  \captionsetup{subrefformat=parens}
  \caption{Examples of branching for an imposed displacement of $\Delta u = \SI{45}{\micro\meter}$. \subref{fig:ex_branch_full} without mass degradation, the crack width is important prior to branching and the exact point at which it happens is hard to identify. \subref{fig:ex_branch_deg} with mass degradation, the width of the crack stays controlled, and the branching is clearly defined.}
  \label{fig:ex_branch}
\end{figure}

When examining branching behaviour, the contrast between the two formulations becomes particularly pronounced.
Two examples of branching are shown in Figures~\ref{fig:ex_branch_full} and~\ref{fig:ex_branch_deg}, corresponding respectively to the cases without and with mass degradation.
In this example, the regularization length is $l_0 = \SI{0.1}{\milli\meter}$ and the imposed displacement is $\Delta u = \SI{45}{\micro\meter}$.

In the absence of mass degradation, the fracture rapidly thickens, and significant instabilities develop before the first two branches finally separate.
It is difficult to clearly characterize the sequence of events preceding branching, and small fragments appear immediately before the formation of the initial pair of branches.
The crack then continues to propagate from one branch, alternately switching between the top and bottom branches as the other is aborted.
A large, highly damaged region forms ahead of the crack tip prior to each branching event.
Although this effect may be less severe for moderate loading levels, it becomes increasingly problematic under extreme loading or fragmentation scenarios, where it is considerably amplified, as illustrated in Figures~\ref{fig:ex_frag_full_0}-\subref{fig:ex_frag_full_3}.

In contrast, the corresponding simulation with mass degradation (Figure~\ref{fig:ex_branch_deg}) exhibits a much more localized process, with the crack width remaining comparatively thin, aside from a slight widening observed before branching.
The branching event itself is considerably more contained, and its initiation point can be identified with substantially greater precision.
It is worth noting that the resulting branching pattern differs markedly from that obtained without mass degradation.
While the physical relevance of this altered pattern lies beyond the scope of the present study, the key observation is that, when fragment identification is required, the mass-degraded formulation provides a more tractable and physically interpretable fracture geometry.

\begin{figure}[ht]
	\centering	
	\includegraphics{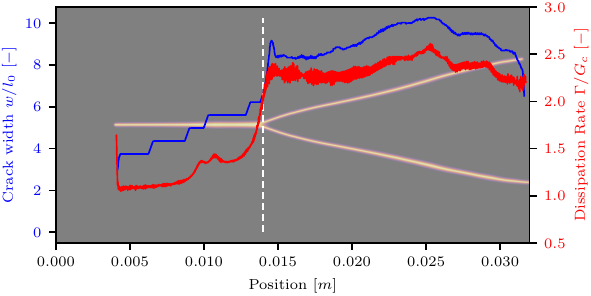}
	\caption{Damage band width (in blue) and dissipated energy (in red) along the crack length. The branching is associated with a damage band width of twice its initial value and a dissipation rate of $2G_c$.}
	\label{fig:damage_width}
\end{figure}

Figure~\ref{fig:damage_width} shows the evolution of the thickness $w$ of the crack along its path, along with the evolution of the dissipation rate $\Gamma$, calculated  with Equations~(\ref{eq:dissipated_energy}) and~(\ref{eq:dissipation_rate}), for the case of mass degradation. The damage width $w$ is computed the same way as in Figure~\ref{fig:damage_width_single}, but with a lower threshold of $d=0.01$ to account for the whole width of the damage profile.
The fracture energy and the crack width typically increase prior to branching~\cite{henry_dynamic_2004, karma_unsteady_2004}, and a critical point is reached when the crack thickness exceeds twice its initial size, i.e. $8l_0$, and when the dissipation rate reaches $2G_c$~\cite{bleyer_dynamic_2017}.
This remains true with mass degradation, but it is much more localized around the point of branching. Initially, the crack thickness is close to $4l_0$, consistently with the analytical prediction.
It then gradually increases along the propagation path, with a marked acceleration shortly before the branching event, at which point the thickness exceeds $8l_0$.
Beyond this threshold, the measured value of $w$ remains high, as the thickness of both branches is included in the computation and no correction is applied for the propagation angle. The dissipation rate $\Gamma$ exhibits a similar structure.
Starting near $G_c$, it increases in two stages: a first rise corresponding to a widening of the crack, followed by a much sharper increase as the fracture begins to branch, ultimately exceeding $2G_c$.

These observations further demonstrate that preserving the wave velocity within the damaged zone is a key factor in limiting the widening of the damage band, thereby yielding more interpretable results in highly dynamic scenarios such as fragmentation.
However, the intention is not to advocate the indiscriminate use of mass degradation as a universal remedy, as this approach introduces its own set of limitations.

\FloatBarrier
\section{Conclusion}
\label{sec:conclusion}

Our simulation results reveal that the interactions between elastic waves and damage play a central role in the formation and growth of damage bands in phase-field modelling of dynamic fracture and fragmentation. In the standard phase-field formulation, stiffness degradation reduces the wave speed in the damaged regions. We have illustrated for an idealized 1D setting that part of the elastic waves are trapped in the damaged zones, which therefore do not behave appropriately as free boundaries. Crucially, elastic waves interacting with the damaged zone can induce additional damage, even when no further energy is supplied. In higher dimensions, for propagating cracks, the standard phase-field formulation leads to thick damage bands that dissipate energy above $G_c$ and propagate at speeds slower than expected. All these problems are alleviated by preserving the wave speed through mass degradation. Our results show that damage bands yield sharper wave reflections and more stable damage behaviour.

Remarkably, our simulation results show encouraging numerical evidence, for a large range of crack velocities, of convergence to Griffith and a well-defined fracture energy. For larger crack velocities, crack branching instabilities develop, and it is not yet clear what triggers those instabilities. Branches obtained with mass degradation are sharper and consistent with previously reported simulations results that show a dependence of branching on crack width and dissipation rate.

There are, however, several limitations to the mass-degradation approach.
First, degrading the mass evidently violates both mass and energy conservation. The approach is also incompatible with tension–compression energy splits: restoring stiffness in compression without restoring mass leads to a wave speed tending toward infinity, and thus to numerical instability in explicit schemes. 
Conversely, restoring the mass in compression also leads to instability, since highly damaged elements may alternate rapidly between tension and compression. 
 
We emphasize that the main benefit of mass degradation is that it yields qualitatively cleaner and more interpretable fragmentation patterns. We have argued that there are some experimental observations suggesting that wave speeds are preserved inside the process zone, but this assumption may not hold for all materials. In cases where the process zone contains microcracks, crushed material, or damaged fibres, the local microstructure may significantly alter wave propagation, and a constant wave speed may no longer be physically justified.

Other strategies could also be explored. For instance, one may adjust the degradation function applied to the mass, seeking a compromise that reduces the crack thickness while retaining some of the mass to limit the increase of wave speed when an energy split is used. Approaches based on plasticity also offer an interesting alternative, particularly when the elastic stiffness is not degraded~\cite{bourdin2025variational}. 
In addition, the yield stress in such models provides a critical stress that is independent of the regularization length, which helps conciliate the choice of a regularization length adapted to the desired resolution and a critical stress corresponding to the material of interest.

\subsection*{Contributions}
\textbf{Shad Durussel:} Conceptualization, Formal analysis, Funding acquisition, Investigation, Methodology, Software, Visualization, Writing – original draft, Writing – review \& editing;
\textbf{Gergely Molnár:} Conceptualization, Methodology, Writing – review \& editing;
\textbf{Jean-François Molinari:} Conceptualization, Funding acquisition, Methodology, Project administration, Supervision, Writing – review \& editing

\subsection*{Acknowledgments}
The authors wish to acknowledge Nicolas Richart and Guillaume Anciaux for their help in the implementation of the phase-field model and optimization solver.
They also thank Laura De Lorenzis for the discussions on phase-field modelling and on implementation of the irreversibility of damage.
This research received founding from the ESA OSIP Idea program, with contract number 4000144496.

\printbibliography

\end{document}